\documentclass[showpacs,preprintnumbers,amsmath,amssymb,prd]{revtex4}
\usepackage{epsfig}
\usepackage{color}
\usepackage{graphicx}
\usepackage{dcolumn}
\usepackage{bm}

\begin{document}
\title{\boldmath Partial Wave Analysis of 
$\chi_{c0}\to\pi^+\pi^-K^+K^-$}
\author{
M.~Ablikim$^{1}$,              J.~Z.~Bai$^{1}$,
Y.~Ban$^{11}$,
J.~G.~Bian$^{1}$,              X.~Cai$^{1}$,
H.~F.~Chen$^{16}$,
H.~S.~Chen$^{1}$,              H.~X.~Chen$^{1}$,
J.~C.~Chen$^{1}$,
Jin~Chen$^{1}$
% ...
\vspace*{2pt}\\(BES Collaboration)\\
}
\affiliation{\vspace*{2pt}
$^1$ Institute of High Energy Physics, Beijing 100049, People's Republic of China \\
% ...
$^{11}$ Peking University, Beijing 100871, People's Republic of China\\
% ...
$^{16}$ University of Science and Technology of China, Hefei 230026, People's Republic of China\\
% ...
}
\vspace*{1pt}
\begin{abstract}
A partial wave analysis of $\chi_{c0}\to\pi^+\pi^-K^+K^-$ in
$\psi(2S)\to\gamma\chi_{c0}$ decay is presented using a sample of
14 million $\psi(2S)$ events accumulated by the BES\,II detector. The data
are fitted to the sum of relativistic covariant
tensor amplitudes for intermediate resonant decay modes. From the
fit, significant contributions to $\chi_{c0}$ decays from
the channels  $f_0(980)f_0(980)$,
$f_0(980)f_0(2200)$, $f_0(1370)f_0(1710)$,
$K^*(892)^0\bar K^*(892)^0$, $K^*_0(1430)\bar K^*_0(1430)$,
$K^*_0(1430)\bar K^*_2(1430) + c.c.$, and $K_1(1270)K$ are found.
Flavor-SU(3)-violating $K_1(1270)-K_1(1400)$ asymmetry is
observed.
Values obtained for the masses and widths of the
resonances $f_0(1710)$, $f_0(2200)$, $f_0(1370)$, and
$K^*_0(1430)$ are presented. 
\end{abstract}

\maketitle

\section{Introduction}

Exclusive heavy quarkonium decays constitute an important laboratory for
investigating perturbative QCD.  Compared to $J/\psi$ decays,
relatively little is known concerning $\chi_{cJ}~(J=0,1,2)$ decays
\cite{PDG}.
More experimental data on exclusive decays of P-wave charmonia
are important for a better understanding of the nature of $\chi_{cJ}$
states, as well as testing QCD based calculations.  Further, the
decays of $\chi_{cJ}$, in particular $\chi_{c0}$ and $\chi_{c2}$,
provide a direct window on glueball dynamics in the $0^{++}$ and
$2^{++}$ channels, as the $\chi_{cJ}$ hadronic decays may proceed via
$c \bar{c} \rightarrow g g \rightarrow q \bar{q} q
\bar{q}$~\cite{close}.

Amplitude analysis of $\chi_{cJ}$ decays is an excellent tool for
studying charmonium decay dynamics. Knowledge of the quantum
mechanical decay amplitude allows one to investigate not only the
intermediate resonant decay modes but also to properly account for the
interference effects between different resonances.

In this paper, partial wave analysis results of
$\chi_{c0}\to\pi^+\pi^-K^+K^-$ in $\psi(2S)\to\gamma\chi_{c0}$ decays
using
14 million $\psi(2S)$ events accumulated at the BES\,II detector are
presented. In previous studies of this channel, only the decay modes
$\chi_{c0}\to K^+\bar K^*(892)^0\pi^- + c.c.$ and $K^*(892)^0\bar
K^*(892)^0$ were measured \cite{mk1,kk}. Here additional information
from partial wave analysis is very important.

\section{Bes detector}

BES\,II is a large
solid-angle magnetic spectrometer that is described in detail in Ref.
\cite{BESII}. Charged particle momenta are determined with a
resolution of $\sigma_p/p = 1.78\%\sqrt{1+p^2}$~($p$ in GeV$/c$) in a
40-layer cylindrical drift chamber. Particle identification (PID) is
accomplished by specific ionization ($dE/dx$) measurements in the
drift chamber and time-of-flight (TOF) measurements in a barrel-like
array of 48 scintillation counters. The $dE/dx$ resolution is
$\sigma_{dE/dx} = 8.0\%$; the TOF resolution is $\sigma_{TOF} = 180$
ps for Bhabha events. Outside of the TOF system is
a 12 radiation length lead-gas barrel shower counter (BSC), operating
in self-quenching streamer mode, that measures the positions and
energies of electrons and photons over 80\% of the total solid angle.
The energy resolution is $\sigma_E/E= 22 \%/\sqrt{E}$ ($E$ in GeV).
Surrounding the BSC is a solenoid magnet that provides a 0.4 T 
magnetic field in the central tracking region of the detector. Three 
double-layer muon
counters instrument the magnet flux return and serve to identify muons
with momentum greater than 0.5 GeV$/c$. They cover $68\%$ of the total
solid angle.

In this analysis, a
 GEANT3 based Monte Carlo (MC) simulation package (SIMBES) with detailed
   consideration of detector performance (such as dead
   electronic channels) is used.
    The consistency between data and MC has been
 checked in
   many high purity physics channels, and the agreement is quite
 reasonable \cite{liuhm}.

\section{Event selection}

The selection criteria described below are similar to those used in
previous BES analyses \cite{BESc,kk}.

\subsection{Photon identification}

A neutral cluster is considered to be a photon candidate when the
angle between the nearest charged track and the cluster is greater
than 15$^{\circ}$, the first hit is in the beginning six radiation
lengths, and the difference between the angle of the cluster
development direction in the BSC and the photon emission direction is
less than 30$^{\circ}$. The photon candidate with the largest energy
deposit in the BSC is treated as the photon radiated from the $\psi(2S)$
and used in a four-constraint (4-C) kinematic fit to the hypothesis
$\psi(2S)\to\gamma\pi^+\pi^-K^+K^-$.

\subsection{Charged particle identification}

Each charged track, reconstructed using MDC information, is
required to be well fit to a three-dimensional helix, be in the polar
angle region $|\cos\theta_{{MDC}}| < 0.80$, and have the point of
closest approach of the track to the beam axis be within 2 cm
of the beam axis and within 20 cm from the center of the interaction
region along the beam line. For each track, the TOF and $dE/dx$
measurements are used to calculate $\chi^2$ values and the
corresponding confidence levels for the hypotheses that the particle
is
a pion, kaon, or proton.

\subsection{Event selection criteria}

Candidate events are
required to satisfy the following selection criteria:

(1) The number of charged tracks is required to be four with net
charge zero.


(2) The sum of the momenta of the two lowest momentum
tracks with opposite charges is required to be greater than 650 MeV$/c$; this removes
contamination from $\psi(2S)\to\pi^+\pi^- J/\psi$ events.

(3) The confidence level for the 4-C kinematic fit to
the decay hypothesis $\psi(2S)\to\gamma\pi^+\pi^-K^+K^-$ is required
to be greater than 0.01.

The combined confidence level determined from the 4-C kinematic
fit and PID information is used to separate
$\gamma\pi^+\pi^-\pi^+\pi^-$, $\gamma K^+ K^- K^+ K^-$, and the
different possible particle assignments for the
$\gamma\pi^+\pi^-K^+K^-$ final states. This combined confidence level  
is defined as $$\int_{\chi^2_{all}}^\infty f(z;ndf_{all})dz,$$
where $f(z;ndf_{all})$ is the $\chi^2$ probability density
function, $\chi^2_{all}$ is
the sum of the $\chi^2$ values from the 4-C kinematic fit
and those of the four track PID assignments,
and $ndf_{all}$ is the corresponding total number of degrees of
freedom. For an event to be selected,
the combined confidence level of $\gamma\pi^+\pi^-K^+K^-$ must be larger than those
of the other possibilities.  In
addition, the PID confidence level of each charged
track must be $>$ 0.01.

Further rejection against $K^0_S\to\pi^+\pi^-$ is obtained by
requiring that any $\pi^+\pi^-K^+K^-$ combination with
$M_{\pi^+\pi^-}$
in the interval ($497\pm50$) MeV/c$^2$ should have $r_{xy} < 5$ mm, where $r_{xy}$ is
the distance from the beam axis to the $\pi^+\pi^-$ vertex.

The invariant mass distribution for the $\pi^+\pi^-K^+K^-$ events
that survive all the above selection requirements is shown in Fig. 1. There
are clear peaks corresponding to the $\chi_{cJ}$ states. The highest
mass peak
corresponds to $\psi(2S)$ decays to four charged track final states that are kinematically
fitted with an unassociated, low energy photon.

\begin{figure}[hbtp]
\begin{center}
\epsfxsize=5.250cm\epsffile{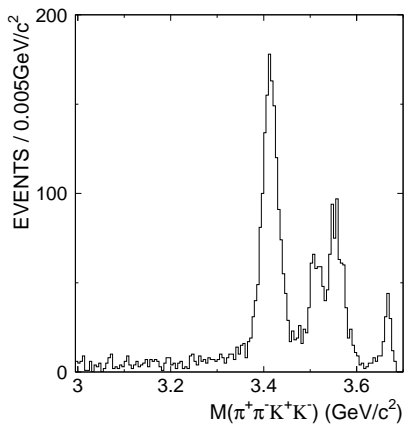}
\label{cpkp}
\vspace*{-5pt}
\caption{The $\pi^+\pi^-K^+K^-$ invariant mass spectrum.  There are
three
  clear $\chi_{cJ}$ peaks.  The highest mass peak corresponds to
  $\psi(2S)$ decays to four charged tracks final states that are kinematically fitted with an
  unassociated, low energy photon. }
\end{center}
\end{figure}

\vspace*{-5pt}
An additional five-constraint (5-C) kinematic fit is made with the
invariant mass of the $\pi^+\pi^-K^+K^-$ being constrained to the
$\chi_{c0}$ mass.  After requiring the confidence level for the 5-C fit
to be greater than 0.01, the $\pi^+\pi^-K^+K^-$
invariant mass distributions for data and MC, shown in Fig.
2, are obtained, where the $\pi^+\pi^-K^+K^-$ invariant mass obtained from
the 4-C fit and the generated mass and width of the $\chi_{c0}$ are
fixed to PDG values \cite{PDG}. The agreement found in this
comparison indicates a clean data sample suitable for partial
wave analysis in which the four-momentum information from the 5-C fit will
be used. Finally, 1371
$\psi(2S)\to\gamma\chi_{c0},~\chi_{c0}\to\pi^+\pi^-K^+K^-$ candidate
events are selected after all the above criteria.

\begin{figure}[hbtp]
\begin{center}
\epsfxsize=4.750cm\epsffile{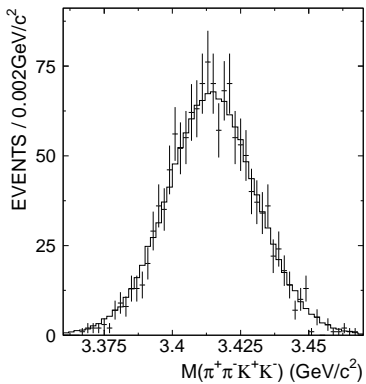}
\label{cpkp}
\vspace*{-2pt}
\caption{$\pi^+\pi^-K^+K^-$ invariant mass distributions from the 4-C
fit after requiring Prob$_{5C} > 0.01$, where the error points are
data and the histogram is MC.}
\end{center}
\end{figure}

\begin{figure}[hbtp]
\begin{center}
\epsfxsize=11.0000500cm\epsffile{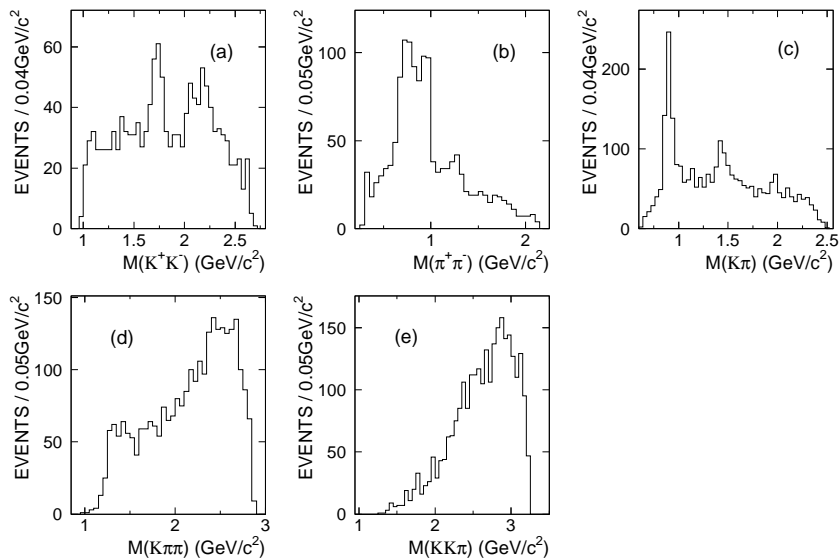}
\label{cpkp}
\vspace*{-2pt}
\caption{The individual (a) $K^+K^-$, (b) $\pi^+\pi^-$, (c) $K\pi$,
  (d) $K\pi\pi$, and (e)
$KK\pi$
invariant mass distributions after application of all selection criteria.}
\end{center}
\end{figure}

Figure 3 shows the individual
$K^+K^-$,~$\pi^+\pi^-$,~$K\pi$,~$K\pi\pi$, and $KK\pi$ invariant mass
distributions.  There are two strong symmetric structures at about
1.75 GeV/c$^2$ and 2.2 GeV/c$^2$ and some evidence for $f_0(980)$ in the $K^+K^-$
mass distribution, and there are clear $\rho(770)$ and $f_0(980)$ peaks and
a smaller one at about 1.3 GeV/c$^2$ in the $\pi^+\pi^-$ mass distribution.
There is no obvious structure for the $KK\pi$
mass spectrum in Fig. 3.

Fig.~4(a) shows the scatter plot of $K^+K^-$ versus $\pi^+\pi^-$
invariant mass for selected
$\psi(2S)\to\gamma\chi_{c0},~\chi_{c0}\to\pi^+\pi^-K^+K^-$ events,
which provides further information on the intermediate resonant decay
modes for $(\pi^+\pi^-)(K^+K^-)$ decay. For instance, it can be seen
that the $f_0(980)\to\pi^+\pi^-$ mainly couples with
$f_0(980),~f_0(1710)$, and $f_0(2200)$, which decay to $K^+K^-$, and
the concentration of events at large $K^+K^-$ mass is also associated with
$\rho(770)$ decays to $\pi^+\pi^-$.

\begin{figure}[htp]
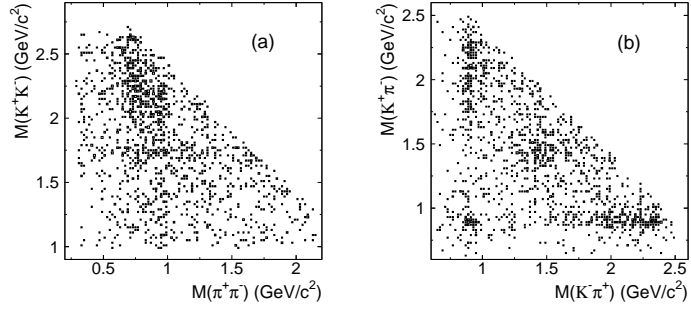

\begin{center}
\epsfxsize=4.1500cm\epsffile{scatter-pipikk.epsi}
\hspace*{0.5cm}
\epsfxsize=4.1500cm\epsffile{scatter-kpikpi.epsi}
\label{cpkp}
\caption{The scatter plots of (a) $K^+K^-$ versus
$\pi^+\pi^-$ and (b) $K^+\pi^-$ versus $K^-\pi^+$ invariant mass for
selected
$\psi(2S)\to\gamma\chi_{c0},~\chi_{c0}\to\pi^+\pi^-K^+K^-$ events.}
\end{center}
\end{figure}

Figure 4(b) shows the scatter plot
of $K^+\pi^-$ versus $K^-\pi^+$ invariant masses. There are obvious clusters
due to $K^*(892)^0$ and around 1.43 GeV/c$^2$,
where there are several known resonances, and evidence for
$(K\pi)$
structures at about 1.7 GeV/c$^2$ and 1.95 GeV/c$^2$, as well as an enhancement above
2 GeV/c$^2$.
There is also a possible 
small accumulation
outside the $K^*(892)^0\bar K^*(892)^0$ cluster, which may be
due to the broad S-wave structure $\kappa$.

The $K\pi\pi$ mass distributions for events where one
$K^{\pm}\pi^{\mp}$ combination is in the $K^*(892)$ mass region $896\pm60$~MeV/c$^2$ and where the $\pi^+\pi^-$ mass in the $\rho(770)$ mass range from 700 to 850
MeV/c$^2$ are shown in Fig. 5. Strong $K_1(1270)$ signals are observed
in both cases, and there is also a weak peak around 1.4 GeV/c$^2$ for
the $K^*\pi$ decay mode. 

\begin{figure}[hbtp]
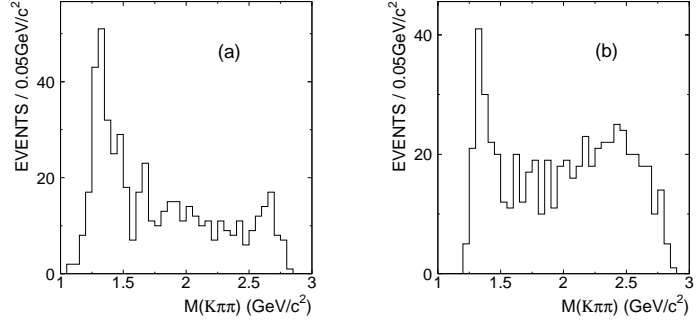

\begin{center}
\epsfxsize=4.05000cm\epsffile{kpipi-kstar.epsi}
\hspace*{0.75cm}
\epsfxsize=4.05000cm\epsffile{kpipi-rho.epsi}
\label{cpkp}
\caption{$K\pi\pi$ mass distributions for events where (a) one
$K^{\pm}\pi^{\mp}$ combination is in the mass region $896\pm60$ MeV/c$^2$ 
and (b) the $\pi^+\pi^-$ mass is in the range 700 - 850 MeV/c$^2$.}
\end{center}
\end{figure}

\section{Analysis method}

We have carried out a partial wave analysis using relativistic
covariant tensor amplitudes constructed from Lorentz-invariant
combinations of the four-vectors and the photon polarization for
$\psi(2S)$ initial states with helicity $\pm 1$ \cite{form1, form2}.  
For $\psi(2S)\to\gamma\chi_{c0},~\chi_{c0}\to\pi^+\pi^-K^+K^-$, the general form 
for the decay amplitude is
\begin {eqnarray*}
A=\psi_\mu(m_1) e^*_\nu(m_2) A^{\mu\nu} =\psi_\mu(m_1)
e^*_\nu(m_2)\sum_i\Lambda_i U_i^{\mu\nu},
\end {eqnarray*}
where $\psi_\mu(m_1)$ is the $\psi(2S)$ polarization four-vector, $e_\nu(m_2)$
is the polarization four-vector of the photon, and 
$U^{\mu\nu}_i$ is the partial wave amplitude with coupling strength determined by a complex parameter $\Lambda_{i}$. For the photon polarization four-vector $e_{\nu}$ with photon momentum $q$, there is the usual Lorentz orthogonality condition $e_{\nu}q^{\nu} = 0$. We assume the Coulomb gauge in the $\psi(2S)$ rest system with momentum $p_{\psi}$, i.e., $e_{\nu}p^{\nu}_{\psi} = 0.$ Then we have 
\begin {eqnarray*}
\sum_{m_2} e^*_\nu(m_2) e_{\nu'}(m_2) = -g_{\nu{\nu'}}+\frac{q_\nu K_{\nu'}+
K_\nu q_{\nu'}}{q\!\cdot\! K}-\frac{K\!\cdot\! K}{(q\!\cdot\! K)^2}q_\nu
q_{\nu'}
\equiv -g^{(\perp\perp)}_{\nu{\nu'}}
\end {eqnarray*}
with $K=p_\psi-q$ and $e_\nu K^\nu=0$. We know that $$\sum^2_{m_1=1}\psi_{\mu}(m_1)\psi^*_{\mu'}(m_1)
=\delta_{\mu\mu'}(\delta_{\mu 1}+\delta_{\mu 2}),$$
so the radiative decay cross
section is:
\begin{eqnarray*}
\frac{d\sigma}{d\Phi}\!&=&\!\frac{1}{2}\sum^2_{m_1=1}\sum^2_{m_2=1}
\psi_{\mu}(m_1) e^*_{\nu}(m_2)A^{\mu\nu}
\psi^*_{\mu'}(m_1) e_{\nu'}(m_2)A^{*\mu'\nu'} \nonumber\\
&=& -\frac{1}{2}\sum_{i,j}\Lambda_i\Lambda_j^*\sum^2_{\mu=1}
U_i^{\mu\nu}g^{(\perp\perp)}_{\nu\nu'}U_j^{*\mu\nu'}
\equiv \sum_{i,j}P_{ij}\cdot F_{ij}
\end{eqnarray*}
where
\begin{eqnarray*}
P_{ij} &= P^*_{ji} &= \Lambda_i\Lambda^*_j, \\
F_{ij} &= F^*_{ji} &= -\frac{1}{2}\sum^2_{\mu=1}
U_i^{\mu\nu}g^{(\perp\perp)}_{\nu\nu'}U_j^{*\mu\nu'}.
\end{eqnarray*}

The partial wave amplitudes $U_i$ for the intermediate states used in the analysis, such as the $K^*(892)^0\bar K^*(892)^0$ are constructed with the four-momenta of the $\pi^+,~\pi^-,~K^+$, and $K^-$, and their specific expressions are 
given in
Ref. \cite{form2}. 
For an intermediate
resonance, the
corresponding Breit-Wigner propagator is denoted by a function:
$$BW={1\over M^2-s-iM\Gamma},$$
where $s$ is the invariant mass-squared and $M$, $\Gamma$ are the
resonance mass and width.
 Angular momenta
$L$ up to 2 in the production process are needed, but higher $L$ give
negligible contributions. Standard Blatt-Weisskopf centrifugal barrier
factors \cite{form1, form2} are included using a radius of interaction 
of 0.8 fm, though results are insensitive to this radius. 

\vspace*{1.5pt}
The relative magnitudes and phases of the
amplitudes are determined by an unbinned maximum likelihood fit. The basis of likelihood fitting is calculating the probability that a hypothesized probability distribution function (PDF) would produce the data set under consideration. If the probability to produce event $i$, characterized by the measurements ${x_i}$, is $P({x_i})$, then the joint probability density for observing the $N$ events in the data sample is $${\cal L} = {\displaystyle \prod_{i=1}^{N}}P({x_i}).$$

The normalization condition for $P({x_i})$ is that its integral over its domain must not depend on the values of the fit parameters. Suppose the differential
cross
section $\displaystyle\bigg({d\sigma \over d\Phi}\bigg)_i$ is the unnormalized PDF for producing event $i$. Then $$P({x_i}) = {\displaystyle\bigg({d\sigma \over d\Phi}\bigg)_i\over \displaystyle\int\bigg({d\sigma \over d\Phi}\bigg) d\Phi},$$ where the integration is over the domain of $\displaystyle\bigg({d\sigma \over d\Phi}\bigg)$.

For the purpose of normalizing the PDF used in fitting the experimental data set, several million MC $\psi(2S)\to\gamma\chi_{c0},~\chi_{c0}\to\pi^+\pi^-K^+K^-$ phase space events were generated.  The events undergo detector simulation and are passed through the same analysis procedure as are the experimental data, and thus the distribution of the MC events passing into the final stage of analysis contains the acceptance information. The normalization integral is then computed as: $$\displaystyle\int\bigg({d\sigma \over d\Phi}\bigg) d\Phi = \sigma\to \displaystyle{1\over N_{MC}}\sum_{i'=1}^{N_{MC}}\displaystyle\bigg({d\sigma \over d\Phi}\bigg)_{i'} = \displaystyle{1\over N_{MC}}\sum_{i'=1}^{N_{MC}}\Bigg({\displaystyle\sum_{j,k}
P_{jk}\cdot F_{jk}\Bigg)_{i'}},$$ where $N_{MC}$ is the number of accepted MC events.

\vspace*{1.5pt}
Background
events obtained from MC simulation 
 are included in the fit, but with the opposite sign of
log likelihood compared to data. These events are used to cancel the backgrounds within
the data sample in the maximum likelihood fit. This technique of background treatment has been used in analyses of Crystal Barrel data (see Ref.~\cite{ANI}) and several previous BES\,II publications \cite{gkk,wpp,phi}.
Then finally we have the logarithm of ${\cal L}$ 
$$\displaystyle \ln {\cal L}= \displaystyle\sum_{i=1}^{N^{'}}\ln\Bigg[{\displaystyle\bigg({d\sigma\over d\Phi}\displaystyle\bigg)_i\over\sigma}\Bigg] 
= 
\displaystyle\sum_{i=1}^{N^{'}}\ln\Bigg[\displaystyle{\displaystyle\bigg(\sum_{j,k}
P_{jk}\cdot F_{jk}\bigg)_i\over \displaystyle{1\over N_{MC}}\displaystyle\sum_{i'=1}^{N_{MC}}\displaystyle\bigg({\displaystyle\sum_{j,k}
P_{jk}\cdot F_{jk}\bigg)_{i'}}}\Bigg],$$
where $N^{'}$ is the total number of data and background events.

For the $\psi(2S)\to\gamma\chi_{c0},~\chi_{c0}\to\pi^+\pi^-K^+K^-$ process, small backgrounds remaining arise mainly from
$\psi(2S)\to\pi^0\pi^+\pi^-K^+K^-$,
$\psi(2S)\to\pi^0\pi^+\pi^-\pi^+\pi^-$,
$\psi(2S)\to\gamma\chi_{c0},~\chi_{c0}\to\pi^+\pi^-\pi^+\pi^-$,
$\psi(2S)\to\gamma\chi_{c1},~\chi_{c1}\to\pi^+\pi^-K^+K^-$, and
$\psi(2S)\to\gamma\chi_{c0},~\chi_{c0}\to K^0_SK^{\pm}\pi^{\mp}$.
 The number of background events is estimated to be about 29, only a few
percent of the data sample, from detailed exclusive and inclusive MC
simulations, and therefore the effect of background is
expected to be minor
in the partial wave analysis.
In the MC simulation we increase the amount of background by a factor of 10 
and then multiply by a normalization
factor of 0.1 to reduce the statistical fluctuation of the background
events.

The optimization of the free parameters $\Lambda_i$ within the amplitude is done using FUMILI \cite{fumili}, which also gives the fitting error matrix. Technically, rather than maximizing $\ln{\cal L}$, this package minimizes $S=-\ln{\cal L}$.  In the minimizing 
procedure, a change in log likelihood of 0.5 represents a
one standard deviation effect.

\section{Analysis results}

Table I shows the decay modes considered in the
partial wave analysis, which are motivated by the structures seen in
the scatter plots of Figs. 4(a) and (b) and in projections of Figs. 3
and 5, changes
$\Delta S$ in log likelihood when the component is dropped from the
fit, and the corresponding statistical significances.  The partial wave amplitude improves log likelihood by more than
5 in most cases. The decay modes $f_0(980)f_0(1710)$, $f_2(1270)f_2(1270)$, and
$f_0(1710)f_0(1370)$, where for the latter $f_0(1710)$ decays to $\pi^+\pi^-$ and $f_0(1370)$ decays to $K^+K^-$, are shown for completeness, but they
are not very significant. The significances 
are calculated from comparing the 
difference between the $S$ values of the fits with and without the component.

\begin{table}[htp]
\begin{center}
\caption{Decay modes fitted in the partial wave analysis, changes
$\Delta S$ in log likelihood when the component is
dropped from the fit, and the corresponding statistical significances.
The errors are statistical only.}
\vspace*{5pt}
\begin{tabular}{l|c|c|r}
\hline
\hline
Decay mode~~~~~~~~&~~~Fitted events~~~&~~~$\Delta S$~~~&~~~Significance\\
\hline
$(\pi^+\pi^-)(K^+K^-)$~~~~~&&\\
\hline
$f_0(980)f_0(980)$&$27.9\pm8.7$&15.7&5.3$\sigma$\\
$f_0(980)f_0(1710)$&$14.7\pm7.0$&5.2&2.8$\sigma$\\
$f_0(980)f_0(2200)$&$77.1\pm13.0$&27.3&7.1$\sigma$\\
$f_0(1370)f_0(980)$&$26.9\pm10.0$&14.6&5.0$\sigma$\\
$f_0(1370)f_0(1710)$&$60.6\pm15.7$&23.5&6.5$\sigma$\\
$f_2(1270)f_2(1270)$&$5.9\pm4.1$&5.8&3.0$\sigma$\\
$\sigma f_0(1710)$&$46.7\pm13.4$&22.2&6.3$\sigma$\\
$\sigma f_0(2200)$&$23.9\pm8.8$&8.5&3.7$\sigma$\\
$f_0(1710)f_0(1370)$&$4.6\pm4.9$&2.5&1.7$\sigma$\\
\hline
$(K^+\pi^-)(K^-\pi^+)$~~~~~&&\\
\hline
$K^*(892)^0\bar K^*(892)^0$&$64.5\pm13.5$&31.1&7.1$\sigma$\\
$K^*(892)^0\bar K^*(1680)^0 + c.c.$&$40.5\pm13.3$&21.0&5.6$\sigma$\\
$K^*_0(1430)\bar K^*_0(1430)$&$82.9\pm18.8$&28.0&7.2$\sigma$\\
$K^*_2(1430)\bar K^*_2(1430)$&$9.2\pm5.3$&7.1&3.3$\sigma$\\
$K^*_0(1430)\bar K^*_2(1430) + c.c.$&$62.0\pm12.1$&40.6&8.7$\sigma$\\
$K^*_0(1430)\bar K^*_0(1950) + c.c.$&$71.0\pm19.1$&22.7&6.4$\sigma$\\
$\kappa\bar\kappa$&$106.8\pm16.7$&39.2&8.6$\sigma$\\
$K^*(892)^0\bar K^*(2300)^0 + c.c.$&$115.7\pm19.4$&45.2&8.8$\sigma$\\
\hline
$(K^{\pm}\pi^+\pi^-)K^{\mp}$&&\\
\hline
$K_1(1270)^+K^- + c.c.$&$153.0\pm19.5$&102.2&13.2$\sigma$\\
$K_1(1400)^+K^- + c.c.$&$19.7\pm8.9$&6.9&2.7$\sigma$\\
$K(1460)^+K^- + c.c.$&$79.7\pm16.8$&39.3&8.2$\sigma$\\
\hline
\hline
\end{tabular}
\end{center}
\end{table}

The statistical uncertainties on the measurements shown in Table I are derived from the uncertainties in the fitting parameters. Recall that the parameters used in the fitting function are magnitudes and
phases of the various processes. The numbers of events are derived from the parameters using numerical integration of the individual amplitudes. Using the numerical expression of the total 
cross section $\sigma$ defined in section IV, the number of fitted events, $N_i$ for an intermediate decay which has one partial wave amplitude $U_i$, is given by the integral over phase space 
(i.e. the sum over the MC events) of the PDF for that decay: $$N_i = 
 \displaystyle{\displaystyle{1\over N_{MC}}\sum_{i'=1}^{N_{MC}}\bigg({\displaystyle P_{ii}\cdot F_{ii}\bigg)_{i'}}\over \sigma}\cdot N^{''},$$ where $N^{''}$ is the number of events after background subtraction. For the intermediate decay with two partial wave amplitudes $U_i,~U_j$, we replace $P_{ii}\cdot F_{ii}$ with 
$P_{ii}\cdot F_{ii} + P_{ij}\cdot F_{ij} + P_{ji}\cdot F_{ji} + P_{jj}\cdot F_{jj}$, where $P_{ij}\cdot F_{ij}$ and $P_{ji}\cdot F_{ji}$ correspond to the interference terms; and those with more amplitudes can be deduced similarly. 

A numerical scheme is used to extract the uncertainties on the numbers of fitted events. The standard deviations and the covariances of the fit parameters are obtained from FUMILI. Next, taking into account the correlation coefficients, a thousand sets of fit parameters are generated using a random number generator CORGEN \cite{V122}. These random numbers are distributed as correlated Gaussian distributions. The number of fitted events for each process is then calculated with each of the thousand sets of parameters, and histogrammed. A Gaussian function is then fit to each of these histograms, and the standard deviations are determined.

The mass projections in
$K^+K^-,~\pi^+\pi^-,~K\pi$, and $K\pi\pi$ are shown in Fig. 6. There
is a reasonable agreement between the data and the fit.

\begin{figure}[hbtp]
\begin{center}
\epsfxsize=8.2499995000cm\epsffile{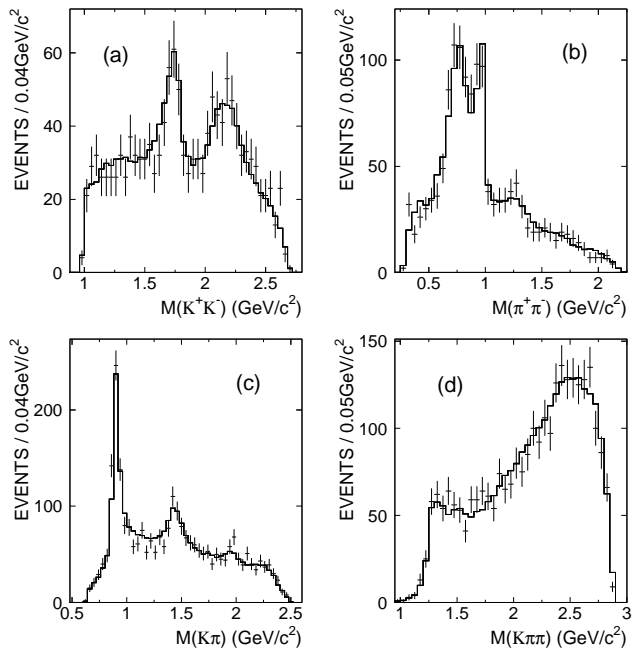}
\label{cpkp}
\vspace*{-2.5pt}
\caption{Mass projections on (a) $K^+K^-$, (b) $\pi^+\pi^-$, (c) $K\pi$ and (d) $K\pi\pi$ for
$\psi(2S)\to\chi_{c0},~\chi_{c0}\to\pi^+\pi^-K^+K^-$. The histograms  represent the fit result, and the points with error bars are data.}
\end{center}
\end{figure}

\subsection{$\chi_{c0}\to(\pi^+\pi^-)(K^+K^-)$}

We begin by discussing the $\chi_{c0}\to (\pi^+\pi^-)(K^+K^-)$ decay
modes. In Fig. 6(b), the $\rho(770)$ due to the decay of
$K_1(1270)\to K\rho(770)$ and a strong $f_0(980)$ are observed. Along the
$f_0(980)~(\to\pi^+\pi^-)$  band in Fig. 4(a), there are
several
enhancements, one at the $K^+K^-$ threshold and others around 1.75 GeV/c$^2$
and 2.2 GeV/c$^2$, 
 which correspond to the
$f_0(980)f_0(980),~f_0(980)f_0(1710)$, and $f_0(980)f_0(2200)$ modes
listed in Table~I.

In the fit, the $f_0(980)$ is described with the usual Flatt\'e
formula \cite {Fla,BS}, and the parameters used are those of Ref.~\cite{BS}.  At higher $\pi^+\pi^-$ mass, a weak signal around 1.3 GeV/c$^2$
is visible.  It is mainly from couplings of the $f_0(1370)$ with
$f_0(980)$ and $f_0(1710)$ which decay to $K^+K^-$. 
A free fit to $f_0(1370)$ gives a fitted mass of $1265 \pm 30$
MeV/c$^2$ and a width of $350 \pm 100$ MeV/c$^2$ with large
statistical errors.  Apart from the above structures, a correlation
between the low $\pi^+\pi^-$ mass enhancement and $K^+K^-$ mass
greater than 1.5 GeV/c$^2$ also appears in Fig.~4(a).  We describe it
by adding $\sigma f_0(1710)$ and $\sigma f_0(2200)$ decay modes, where
the parameterization of a $J/\psi\to\omega\pi^+\pi^-$ analysis (Adler
zero parameterization)~\cite{wpp} is adopted to describe the broad
S-wave $\sigma$. The $\sigma f_0(1710)$ and $f_0(1370)f_0(1710)$
decays dominate the production of $f_0(1710)$ in the partial wave fit
of $\psi(2S)\to\gamma\chi_{c0},~\chi_{c0}\to\pi^+\pi^-K^+K^-$.

The fitted mass and width of the $f_0(1710)$ are M $=1760\pm15$ MeV/c$^2$
and $\Gamma = 125\pm 25$~MeV/c$^2$. The mass is somewhat higher than the PDG
value \cite{PDG}. The $f_0(1790)$ has been  recently claimed in $J/\psi$
decays \cite{phi}. The parameters of $f_0(1790)$ and those of
$f_0(1710)$
given by the PDG and Ref. \cite{gkk} are tested in this analysis, 
and the log
likelihood becomes worse by 10.3, 11.5, and 4.2, respectively. The log
likelihood
can be improved by 5.7 if both the $f_0(1790)$ and $f_0(1710)$ (using
parameters of Ref. \cite{gkk}) are
included into the fit, while the results for the other components
only change a little.
A fit replacing the decay modes with an $f_0(1710)$
with $f_2(1710)$ -- namely, using  $\sigma
f_2(1710),~f_0(980)f_2(1710)$, and $f_0(1370)f_2(1710)$ decays
instead of $\sigma f_0(1710),~f_0(980)f_0(1710)$, and
$f_0(1370)f_0(1710)$ in the fit -- is made in order to check the
spin-parity of the structure around 1.75 GeV/c$^2$ in the $K^+K^-$ mass
region. It gives a worse log likelihood by 55.6 and a poor fit
result, as shown in Fig. 7(a).

\begin{figure}[hbtp]
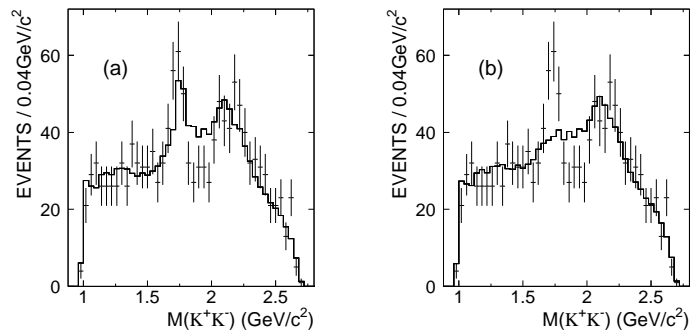

\begin{center}
\epsfxsize=4.025000cm\epsffile{proj-f21710.epsi}
\hspace*{0.75cm}
\epsfxsize=4.025000cm\epsffile{proj-rhop.epsi}
\label{cpkp}
\vspace*{-5pt}
\caption{Mass projections in $K^+K^-$ for fits
replacing $f_0(1710)$ with (a) $f_2(1710)$ or (b) $\rho(1450)$ and
$\rho(1700)$. The dots with error  bars are data, and the histograms
are the fits.}
\end{center}
\end{figure}

There are two
$\rho$-like resonances, $\rho(1450)$ and $\rho(1700)$, which decay to
$K\bar
K$ final states in the 1600-MeV/c$^2$ region~\cite{PDG}. As a check, we add
$\rho(770)\rho(1450)$ and $\rho(770)\rho(1700)$ intermediate decay
modes into
the fit with four more fitted parameters, using the mass
and width\
 measurements given by
Ref. \cite{CBAR}, and the log likelihood is improved by
$6.4$. Replacing
the components $\sigma f_0(1710)$, $f_0(980)f_0(1710)$, and 
$f_0(1370)f_0(1710)$
with $\rho(770)\rho(1450)$ and $\rho(770)\rho(1700)$ gives a worse log
likelihood by 81.6, and
the corresponding fit is shown in Fig.~7(b). An investigation of the $\rho(770)$ band in Fig. 4(a) indicates that the structures at high $K^+K^-$ mass are a reflection of the other intermediate decay modes. 

A free fit to $f_0(2200)$ gives a fitted mass of $2170 \pm 20$ MeV/c$^2$ and
a width of $220 \pm 60$ MeV/c$^2$. 
There are several scalar and tensor resonances located in the mass
range greater than 2.0 GeV/c$^2$ which decay to the $K^+K^-$ final states~\cite{PDG}: $\rho(2150)$, $f_2(2150)$, $f_0(2200)$, and $f_2(2300)$.  A
variety of alternative fits to the high $K^+K^-$ mass region, listed
in Table~II, using these resonances and the $f_0(2100)$
have been tried.
Note that PDG mass
and width values~\cite{PDG} are used in the fits for the $f_0(2100)$,
$f_2(2150)$, $\rho(2150)$, and $f_2(2300)$ and that $f_0(2100)$,
$f_2(2150)$, $f_0(2200)$, and $f_2(2300)$ couple with the $\sigma$ and
$f_0(980)$, while the $\rho(2150)$ couples with the $\rho(770)$.
We also tried $f_2(2200)$ instead of the $f_0(2200)$ in one fit, assuming it couples with the $\sigma$ and
$f_0(980)$.

\begin{table}[htp]
\begin {center}
\caption {Alternative fits to the high $K^+K^-$ mass region instead of using only a $f_0(2200)$ involved in the fit; the right-hand
column shows
values of $\Delta S$; positive values indicate poorer fits.}
\vspace*{5pt}
\begin {tabular} {lr}
\hline
\hline
Components &  ~~~~~$\Delta S$\\
\hline
a)~~$\rho(2150)$&+35.3\\
b)~~$f_2(2200)$~\,~M$ = 2170$ MeV$/c^2$,&+33.1\\
~~~~~~~~~~~~~~~~\,~~$\Gamma = 220$ MeV$/c^2$&\\
c)~~$f_2(2150) + f_0(2200)$&-4.5\\
d)~~$f_2(2150) + f_0(2100)$&+9.9\\
e)~~$f_2(2150) + f_2(2300)$&+28.6\\
f)~~$f_0(2100) + f_0(2200)$&-4.3\\
g)~~$f_0(2100) + f_2(2300)$&-3.5\\
h)~~$f_0(2200) + f_2(2300)$&-9.0\\
i)~~$f_2(2150)$&~~~~~+32.5\\
\hline
\hline
\end {tabular}
\end {center}
\end {table}

In Table II, fit a), using $\rho(770)\rho(2150)$ instead of $\sigma
f_2(2200)$ and $f_0(980)f_0(2200)$, gives a worse log likelihood by
35.3. A fit with both $f_0(2200)$ and $\rho(2150)$
improves the log likelihood by about 3.0 compared to the case
with $f_0(2200)$ only. Also b), e), and i),
where no scalar is included, as well as d), give bad fits.
Fits c), f), g), and h) improve the log likelihood a little, but not
significantly. 

Based on these fit results, we conclude that a scalar state $f_0(2200)$ 
at about 2.2 GeV/c$^2$ which decays to $K^+K^-$ is needed in this
channel, but no additional significant $0^{++}$ or
$2^{++}$
is required in this mass region according to our study in Table II.

\subsection{$\chi_{c0}\to(K^+\pi^-)(K^-\pi^+)$}

The shape of the $K^*(892)$ is described by a P-wave relativistic
Breit-Wigner curve, with a width $$\Gamma = \Gamma_0
{{m_0}\over{m}}{{1+r^2p_0^2}\over{1+r^2p^2}}\bigg\lbrack{{p}\over{p_0}}\bigg
\rbrack^3,$$
where $m$ is the mass of the $K\pi$ system, $p$ is the momentum of
kaon in the $K\pi$ system, $\Gamma_0$ is the width of the
resonance, $m_0$ is the mass of the resonance, $p_0$ is $p$ evaluated at the resonance mass, $r$ is the
interaction radius and $\displaystyle
{{1+r^2p_0^2}\over{1+r^2p^2}}$ represents
the contribution of the barrier factor.
The
value $(3.4\pm0.6\pm0.3)$ (GeV$/c$)$^{-1}$ measured by the $K^-\pi^+$
scattering experiment \cite{ASTON} as
an approximate estimation of the interaction radius $r$ is used.

Adding a
$K^*(892)^0\bar K^*(1680)^0$ decay mode improves log likelihood by 21.0, where
the mass and width of $K^*(1680)^0$ are fixed to PDG values \cite{PDG}
and the interaction radius $r$ in the fit is set to the value 2.0
given by
Ref. \cite{ASTON}.

Most of the peak at 1430 MeV/c$^2$ is fitted as $K^*_0(1430)\bar
K^*_0(1430)$ and $K^*_0(1430)\bar
K^*_2(1430)$
$+~c.c.$, with only a small contribution from
$K^*_2(1430)\bar
K^*_2(1430)$ -- the fitted number of $K^*_2(1430)\bar
K^*_2(1430)$ events is about 11\% of that of $K^*_0(1430)\bar
K^*_0(1430)$. Adding the $K^*_2(1430)\bar
K^*_2(1430)$ mode improves the log likelihood by 7.1, which
corresponds
to about a significance of 3.3$\sigma$ for two free parameters of the
fitted amplitudes. The mass and width of the $K^*_2(1430)$
are fixed to the PDG values \cite{PDG}.  A free fit to
$K^*_0(1430)$ gives a fitted mass of $1455 \pm 20$ MeV/c$^2$ and
a width of $270 \pm 45$~MeV/c$^2$.

There is another structure
visible around 1.95 GeV/c$^2$ in the $K\pi$ mass distribution.
 Among the three generalized C-parity
allowed intermediate decays from $\chi_{c0}$, 
$K^*_0(1430)\bar K^*_0(1950)$, $K^*_2(1430)\bar K^*_0(1950)$, and
$\kappa\bar K^*_0(1950)$, the first decay gives a better
log likelihood value in our fit than the other two by 22.9 and 14.6,
respectively.
 A free fit gives a fitted mass of $1945\pm
30$ MeV/c$^2$, but the width is poorly determined, $\sim 500$ MeV/c$^2$, so we
adopt the PDG mass and width values for the $K^*_0(1950)$ in
the partial wave analysis.

In Fig. 4(b) there is a  possible accumulation 
outside
the $K^*(892)^0\bar K^*(892)^0$ cluster. Its explanation may be the
broad S-wave
$\kappa$. Since the properties of $\kappa$ are still 
controversial (cf., the review paper in the
PDG \cite{PDG}, and references therein), we use a
Breit-Wigner amplitude of
constant width, without any phase space factor with the
parameterization obtained in Ref. \cite{kappa1}, M$= 790$ MeV/c$^2$, $\Gamma =
860$ MeV/c$^2$ to describe this broad structure. The log
likelihood becomes worse by 39.2 after removing the
$\kappa\bar
\kappa$ from the fit,  and the corresponding projection in $K\pi$ mass,
shown in Fig. 8, obviously disagrees with data at low $K\pi$ mass.

\begin{figure}[hbtp]
\begin{center}
\epsfxsize=4.750000cm\epsffile{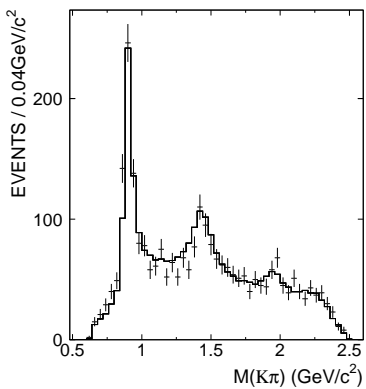}
\label{cpkp}
\vspace*{-5pt}
\caption{The  $K\pi$ mass projection for the fit
without $\kappa\bar \kappa$ decay.}
\end{center}
\vspace*{-10pt}
\end{figure}


Finally, an intermediate decay mode $K^*(892)^0\bar
K^*(2300)^0$ is added into the fit. Here the $K^*(2300)^0$ is used as an effective structure
to describe the poorly known high $K\pi$ mass region.
The log likelihood changes by 45.2 if the $K^*(892)^0\bar
K^*(2300)^0$ is omitted.  The scanned mass and width for
$K^*(2300)^0$ are $\sim 2.3$ GeV/c$^2$ and 300 MeV/c$^2$
respectively.  No
higher spin-parity tests such as $3^-$ or $5^-$ are made.

\subsection{$\chi_{c0}\to (K^{\pm}\pi^+\pi^-)K^{\mp}$}

In Fig. 5 there are strong
$K_1(1270)$ signals in both $K\rho(770)$ and $K^*(892)\pi$,
and there is also a
weak peak around 1.4~GeV/c$^2$ in the $K^*(892)\pi$. In the partial
wave analysis, $K_1(1270)K$ and $K_1(1400)K$ decays, where
$K_1(1270)$ decays to $K\rho(770),~K^*(892)\pi$, and $K^*_0(1430)\pi$ 
and $K_1(1400)$ decays to $K^*(892)\pi$, are added. The fit also requires an 
additional structure to describe the $K\pi\pi$ mass distribution at about
1.4~GeV/c$^2$. Here adding $K(1460)$ in the fit will improve the log
likelihood by
 39.3; in the fit, the mass and width values given by the PDG \cite{PDG} are used. The masses and widths of the $K_1(1270)$
and $K_1(1400)$ are also fixed to PDG values so as to reduce
uncertainties.

There are two lowest-lying axial-vector-meson octets. These correspond
to the singlet $(^1P_1)$ and triplet $(^3P_1)$ spin configurations of
two quarks in a P-wave orbital angular momentum state. The
nonstrange, isospin $I = 1$ members of the two octets have opposite
$G$ parity: the $b_1(1235)$ is in the $^1P_1$ octet and has $G = +1$,
while the $a_1(1260)$ is in the $^3P_1$ octet and has $G = -1$. 
 The strange members of the
$^3P_1$ and $^1P_1$ octets, the $K_A$ and $K_B$, respectively, are
mixtures of the observed physical states, the $K_1(1270)$ and the
$K_1(1400)$, where  
\begin {eqnarray*} K_A = \cos\theta K_1(1400) + \sin\theta
K_1(1270),\\
K_B = \cos\theta K_1(1270) - \sin\theta
K_1(1400),
\end {eqnarray*}
and the mixing angle is near $\theta\simeq 45^{\circ}$
\cite{theta}. The dominant $K_1(1270)$ decay mode is to $K\rho(770)~(
{\cal{B}} = 42 \%\pm6 \%)$; the $K_1(1400)$ decays almost always to
$K^*(892)\pi~({\cal {B}} = 94 \%\pm 6 \%).$ In the limit of
strict flavor SU(3) symmetry, the amplitudes for two-body decays to
conjugate mesons in the same pair of octets should be equal. Thus,
since decays to $b_1\pi$ are forbidden by $G$ parity, decays to
$K_B\bar K$ are disallowed by SU(3), and one expects relatively pure
$K_A\bar K$ final states in $\chi_{c0}$ decays. And, since
$\theta\simeq 45^{\circ}$, there should be roughly equal amounts of
$K_1(1270)$ and $K_1(1400)$.

The remarkable feature of the $K\pi\pi$ distribution is that the
contribution of the $K_1(1400)$ in the final fit is very small
compared to the large $K_1(1270)$ signal. Adding a $K_1(1400)$ improves
the log likelihood by 6.9 which corresponds to the significance of
about 2.7$\sigma$. The partial wave analysis yields a
$K_1(1270)\to K\rho(770)$ signal of $68.3\pm13.4$ events and a
$K_1(1400)\to K^*(892)\pi$ signal of $19.7\pm 8.9$ events. The function
used to fit the $K_1(1270)$ is a convolution of an S-wave Breit-Wigner
function for
the
$K_1(1270)$ with a P-wave Breit-Wigner function for the $\rho(770)$ meson.
Flavor-SU(3)-violating $K_1(1270)-K_1(1400)$ asymmetry for
$\chi_{c0}$ decay is observed.

\subsection{GOODNESS-OF-FIT}

To determine
the
goodness of fit, a
$\chi^2$ is calculated by comparing  histograms, $e.g.,$ a
vector of Poisson distributed numbers ${\bf n} = (n_1,...,n_N),$ with
a hypothesis for their expectation values $\nu_i = E[n_i]$. As the
distribution is Poisson with variances $\sigma_i^2 = \nu_i$, the
$\chi^2$ becomes $Pearson's~\chi^2~statistic$, $$\chi^2 =
\sum_{i=1}^{N} \frac{(n_i-\nu_i)^2}{\nu_i}.$$ If the hypothesis ${\bf
\nu} = (\nu_1,...\nu_N)$ is correct and if the measured values $n_i$
are sufficiently large, then the $\chi^2$ statistic will follow the
$\chi^2$ probability density function with the number of degrees of freedom (ndf) equal to
the
number of measurements $N$ minus the number of fitted parameters.

For an $n$-body final state, the number of independent kinematic
variables is $3n -4$. Thus, one can compare $3n -4$ selected
independent
variables
with the fit results by defining a quantity $$\chi^2_{all} =
\sum_{j=1}^{3n-4}\chi_j^2$$ to check the quality of a global fit,
which obeys the $\chi^2$ distribution approximately with the number of
degrees of freedom equal to the
total number of measurements minus the number of fitted parameters;
and
the individual $\chi^2_j$ will give a qualitative measure of the
goodness of fit for each
kinematic variable.

The number of independent kinematic variables for
$\psi(2S)\to\gamma\chi_{c0},$~$\chi_{c0}\to
\pi^+\pi^-K^+K^-$ process
is 10 after the use of the 5-C fit with the additional $\chi_{c0}$
mass constraint. We use the following 10 distributions to
check
the goodness-of-fit:

\vspace*{1pt}
--~$(\theta,~\phi)_{\pi\pi KK}$~=~the polar angle and azimuthal angle
  of
the ($\pi^+\pi^-K^+K^-$) in the laboratory system,

\vspace*{1pt}--~$(\theta,~\phi)_{\pi\pi K}$~=~the polar angle and azimuthal
  angle of
the ($\pi^+\pi^-K^-$) in the ($\pi^+\pi^-K^+K^-$) center of mass,

\vspace*{1pt}--~$(\theta,~\phi)_{\pi K}$~=~the polar angle and azimuthal
  angle of
the ($\pi^+K^-$) in the ($\pi^+\pi^-K^-$) center of mass,

\vspace*{1pt}--~$(\theta,~\phi)_{\pi}$~=~the polar angle and azimuthal
  angle of
the $\pi^+$ in the ($\pi^+K^-$) center of mass,\\
and the invariant masses for the $({\pi^+\pi^-K^-})$ and $({\pi^+K^-})$
systems. Table III shows the results, where the number of bins
is taken  as the number of degrees of freedom for each individual
distribution,
and Fig.~9 shows the projections of the 10
 variables. 
There is
 excellent agreement
between the data and fit.

\begin{table}[htp]
\begin {center}
\caption {Check of goodness of fit using 10 independent kinematic
variables, where ndf and C.L. are the number of degrees of freedom and the corresponding confidence level.}
\vspace*{5pt}
\begin {tabular} {lcccr}
\hline
\hline
Variable&$\chi^2$&ndf&$\chi^2$/ndf&~~~C.L.\\
\hline
$cos\theta_{\pi\pi KK}$~~~&~~~24.46~~~&~~~18~~~&~~~1.36~~~&0.14\\
$\phi_{\pi\pi KK}$~~~&~~~14.23~~~&~~~20~~~&~~~0.71~~~&0.82\\
$cos\theta_{\pi\pi K}$~~~&~~~11.29~~~&~~~20~~~&~~~0.56~~~&0.94\\
$\phi_{\pi\pi K}$~~~&~~~22.58~~~&~~~20~~~&~~~1.13~~~&0.31\\
$cos\theta_{\pi K}$~~~&~~~11.34~~~&~~~16~~~&~~~0.71~~~&0.79\\
$\phi_{\pi K}$~~~&~~~18.81~~~&~~~20~~~&~~~0.94~~~&0.53\\
$cos\theta_{\pi}$~~~&~~~12.15~~~&~~~20~~~&~~~0.61~~~&0.91\\
$\phi_{\pi}$~~~&~~~12.63~~~&~~~20~~~&~~~0.63~~~&0.89\\
$M_{\pi^+\pi^- K^-}$~~~&~~~37.26~~~&~~~36~~~&~~~1.04~~~&0.41\\
$M_{\pi^+ K^-}$~~~&~~~55.07~~~&~~~47~~~&~~~1.17~~~&0.20\\
\hline
\hline
\end {tabular}
\end {center}
\end {table}

\begin{figure}[hbtp]
\begin{center}
\epsfxsize=17.50cm\epsffile{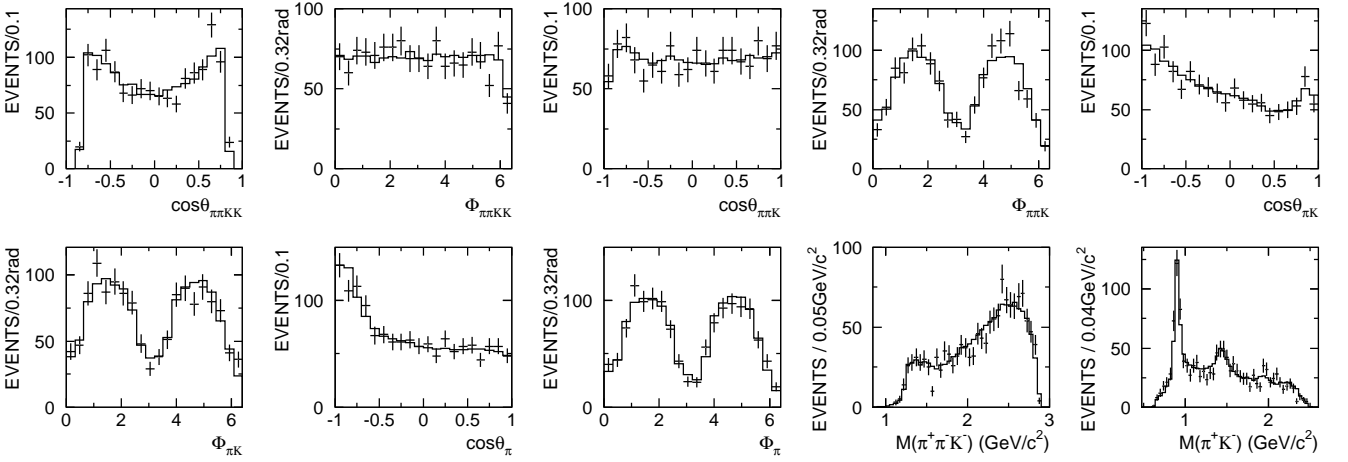}
\label{cpkp}
\caption{Fit projections for the 10 variables described in the text after the global fit.}
\end{center}
\end{figure}

Using the results in Table III, the $\chi_{all}^2$ obtained is 219.8
with 185 degrees of freedom (=237-52, where 237 is a sum of the bin
numbers for the 10 distributions and 52 is the number of fitted
amplitude parameters), which corresponds to a confidence level of 4\%.

\section{Systematic error}

In this analysis, the systematic errors are estimated by considering
the following sources:

(1) Uncertainty of the parameterization of the $\sigma$ line shape:
use a
Breit-Wigner amplitude of constant width M = 470 MeV/c$^2$, $\Gamma = 613$
MeV/c$^2$ \cite{wpp} instead of the Adler zero  parameterization. 

(2) Uncertainty of the parameterization of the $\kappa$ line shape: change the mass
and width of the Breit-Wigner amplitude of constant width to M = 745
MeV/c$^2$, $\Gamma = 622$ MeV/c$^2$ \cite{kappa1}.

(3) The parameters of the $f_0(980)$ are still uncertain, and in addition to the solution of
Ref. \cite{BS}, we also use 
measurements of some recent
experiments such as E791, GAMS, and WA102 \cite{E791,BELL,WA102}, where a
Breit-Wigner description with the width varying from 44
to 80 MeV/c$^2$ was used for the $f_0(980)$.
We determine the change both by using
the solutions of Refs. \cite{E791,BELL,WA102} and by varying $g_1$ in
Ref. \cite{BS} from
0.1108 GeV/c$^2$ to 0.090 GeV/c$^2$ and 0.130 GeV/c$^2$ while keeping the ratio
$g_2/g_1$
fixed.

(4) As mentioned above, we use the measurement of Ref.~\cite{ASTON},
$(3.4\pm0.6\pm0.3)$ (GeV$/c$)$^{-1}$ for $r$ in the P-wave
relativistic
Breit-Wigner parameterization. We also use $r$ varied by one sigma
to 2.73 (GeV$/c$)$^{-1}$ and 4.07
(GeV$/c$)$^{-1}$
to determine the change in the fit.

(5) Vary mass and width values of the $(K\pi)$ components
    $K^*_2(1430)$,
$K^*(1680)$, and $K^*_0(1950)$ within
the PDG
   errors~\cite{PDG}.

(6) Vary mass and width values of the $(K\pi\pi)$ components
    $K_1(1270)$,
$K_1(1400)$, and $K(1460)$ within
the PDG
   errors~\cite{PDG}, where the width of $K(1460)$ is changed to 200
 and 300 MeV/c$^2$  and mass
to 1.40 and 1.46 GeV/c$^2$.

(7) Remove the small component $f_0(1710)f_0(1370)$ from the fit, where 
$f_0(1710)$ decays to $\pi^+\pi^-$ and $f_0(1370)$ 
decays to $K^+K^-$.

(8) Add three $(K\pi\pi)$ resonances $K_1(1650),~K_2(1770)$, and
$K_2(1820)$ into the fit. 

(9) Remove the $K^*(892)^0\bar K^*(2300)^0 + c.c.$ decay mode
from the original fit but keep (8). Tests (8) and (9) strongly affect the
branching fraction measurements. 

(10) Try alternative starting conditions for the maximum
likelihood fit. 

(11) Uncertainty of the background in the partial wave analysis fit. In order
to investigate the effect of the amount and shape of background on the results, we increased the number of background events in the
fit to 55, which is the maximum estimation, fitted without background, and  
added an incoherent phase space background normalized to 29 events instead of fitting with the opposite sign of log likelihood for the MC background.
The change on the fit results is very small, less than 5\% for all the
measurements.  

Total errors are obtained by adding the individual errors in
quadrature. For the branching fraction uncertainties, the
uncertainties in MDC tracking, kinematic fitting, PID, efficiency of
the photon ID, and the number of $\psi(2S)$ events \cite{moxh} are
also included, and the total systematic error for this part, 12\%, is
taken from Ref. \cite{kk}.

\section{DISCUSSION}

From the $\chi_{c0}\to\pi^+\pi^- K^+K^-$ decay fit results, it is
found that scalar resonances have large decay fractions compared to
those of tensors, and such decays provide a relatively clean
laboratory in which to study the properties of the scalars
$f_0(980),~f_0(1370),~f_0(1710)$, and $f_0(2200)$. There is also conspicuous
production due to the $K^*(892)^0\bar K^*(892)^0$ and $K^*_0(1430)\bar
K^*_0(1430)$ (and $K^*_0(1430)\bar K^*_2(1430) + c.c.$) pairs, and
flavor-SU(3)-violating $K_1(1270)-K_1(1400)$ asymmetry is
observed. Information on these states is very desirable and will be
described below. 

For other components in Table I, because of the low statistics
($f_2(1270)f_2(1270)$, $f_0(1710)f_0(1370)$, and $K^*_2(1430)\bar
K^*_2(1430)$), uncertain parameters of intermediate resonances involved
($\sigma f_0(1710)$, $\sigma f_0(2200)$, $K^*(892)^0\bar K^*(1680)^0 +
c.c.$, $K^*_0(1430)\bar K^*_0(1950) + c.c.$, $\kappa\bar\kappa$, and
$K(1460)K$), and the poorly known high mass $K\pi$ region which is
described by the $K^*(892)^0\bar K^*(2300)^0 + c.c.$, it
is difficult to obtain precise quantitative results or make definite
conclusions.
The systematic errors on the numbers of
events for decay modes $f_0(980)f_0(1710)$ and $f_0(1370)f_0(980)$ are
very large, while the significance of the $f_0(980)f_0(1710)$ is
small; therefore we also will not consider these two decay modes in the
following branching fraction measurements.

The numbers used in the branching fraction (or upper limit)
calculations
and the corresponding results are summarized in Table IV, where the first
errors are statistical and the
second are systematic. The
value of
${\cal B}[\psi(2S)\to\gamma\chi_{c0}]$, $(9.22\pm0.11\pm0.46)\%,$
recently measured by
the CLEO experiment, is used \cite{cleoc}. The fit removing the
$K_1(1270)\to K\rho(770)$
gives a worse log likelihood by 39.5, which corresponds to the
signal significance of about
$8.6\sigma$. The $90\%$ confidence level (C.L.) upper limit for $\chi_{c0}\to
K_1(1400)\bar{K}$ is obtained by increasing the
number of events by $1.28\sigma$, where $\sigma$ includes the
statistical and the systematic errors added in
quadrature.

\begin{table}[htp]
\begin {center}
\caption {Summary of numbers used in the branching fraction (or upper
limit)
calculations 
and corresponding results, where $X$ represents the intermediate decay modes, $N^{fit}$ is the number of fitted events, and $\epsilon$ is the efficiency.}
\begin {tabular} {l|ccccr}
\hline
\hline
Decay mode&~~~~~~$N^{fit}$~~~~~~&~~~$\epsilon~(\%)$~~~&~~Sys. error~(\%)~~~&${\cal
B}[\chi_{c0}\
\to
X\to\pi^+\pi^- K^+K^-]$&~~~Significance\\
&&&&$(\times 10^{-4})$&\\
\hline
${f_0(980)f_0(980)}$&$27.9\pm8.7$&$6.25\pm0.01$&$^{+55.7}_{-45.3}$&$3.46\pm
1.08^{+1.93}_{-1.57}$&$5.3\sigma$\\
${f_0(980)f_0(2200)}$&$77.1\pm13.0$&$7.09\pm0.01$&$^{+19.6}_{-27.2}$
&$8.42\pm1.42^{+1.65}_{-2.29}$&$7.1\sigma$\\
${f_0(1370)f_0(1710)}$&$60.6\pm15.7$&$6.59\pm0.01$&$^{+46.1}_{-23.6}$&$7.12
\pm1.85^{+3.28}_{-1.68}$&$6.5\sigma$\\
${K^*(892)^0\bar
K^*(892)^0}$&$64.5\pm13.5$&$6.18\pm0.01$&$^{+28.3}_{-24.6}$&$8.09\pm1.69^{+2.29}_{-1.99}$&$7.1\sigma$\\
${K^*_0(1430)\bar
K^*_0(1430)}$&$82.9\pm18.8$&$6.15\pm0.01$&$^{+29.2}_{-18.2}$&$10.44\pm2.37^{+3.05}_{-1.90}$&$7.2\sigma$\\
${K^*_0(1430)\bar K^*_2(1430)} +
c.c.$&$62.0\pm12.1$&$5.66\pm0.01$&$^{+15.6}_{-23.4}$&$8.49\pm1.66^{+1.32}_{-1.99}$&$8.7\sigma$\\
${K_1(1270)^{+}K^{-} + c.c.,}$&&&&&\\
~~~$K_1(1270)\to K\rho(770)$&$68.3\pm13.4$&$5.68\pm0.01$&$^{+19.4}_{-17.6}$&$9.32\pm1.83^{+1.81}_{-1.64}$&$8.6\sigma$\\
${K_1(1400)^{+}K^{-} + c.c.,}$&&&&&\\
~~~$K_1(1400)\to K^*(892)\pi$&$19.7\pm8.9$&$4.94\pm0.01$&$^{+219}_{-24.5}$&$< 11.9$ (90\%
C.L.) &$2.7\sigma$\\
\hline
\hline
\end {tabular}
\end {center}
\end {table}

The partial wave fit provides magnitudes and phases of the different
partial amplitudes, as well as the interference terms. The intensity
from these amplitudes is used to weight both the complete set of
generated MC events and the set which survives the selection
procedure. The ratio between these two weighted sets is the
efficiency.

Using the number of selected
$\psi(2S)\to\gamma\chi_{c0}\to\gamma\pi^+\pi^-K^+K^-$ events, the
overall efficiency determined by the method above, $(5.85\pm0.01)\%$,
and the result of Ref. \cite{cleoc} we
get
the corresponding branching
fractions after subtracting background \begin {eqnarray*} {\cal
B}[\psi(2S)\to\gamma\chi_{c0}\to\gamma\pi^+\pi^-K^+K^-] =
(1.64\pm0.05\pm0.20)\times 10^{-3},\\{\cal
B}[\chi_{c0}\to\pi^+\pi^-K^+K^-] = (1.78\pm0.05\pm0.23)\times 10^{-2}.\end{eqnarray*}

For the decay mode $\chi_{c0}\to f_0(980)f_0(980)$,
each $f_0(980)$ can decay to either $\pi^+\pi^-$ or $K^+K^-$, so it is
necessary to divide the result in Table IV
by a factor of 2 to obtain 
the branching
fractions
\begin {eqnarray*} {\cal
B}[\psi(2S)\to\gamma\chi_{c0}\to\gamma
f_0(980)f_0(980)]{\cal B}[f(980)\to\pi^+\pi^-]{\cal B}[f(980)\to K^+K^-] =
(1.59\pm0.50^{+0.89}_{-0.72})\times 10^{-5},\\
{\cal B}[\chi_{c0}\to
f_0(980)f_0(980)]{\cal B}[f(980)\to\pi^+\pi^-]{\cal B}[f(980)\to K^+K^-] =
(1.73\pm0.54^{+0.96}_{-0.78})\times 10^{-4}.
\end{eqnarray*} 
Combining this result and that
of Ref. \cite{980},
we can determine
the ratio of the partial decay
width of $f_0(980)$ to $\pi\pi$ to those of  $\pi\pi$ and $K\bar K$:
$${{\Gamma_{\pi\pi}}\over{\Gamma_{\pi\pi}+\Gamma_{K\bar K}}} =
{{(6.5\pm1.93)\times {{3}\over{2}}}\over{(6.5\pm1.93)\times
{{3}\over{2}}}+ (1.59^{+1.00}_{-0.86})\times
2 } = 0.75^{+0.11}_{-0.13}.$$
The numerical factors ${{3}\over{2}}$ and 2 take into account
that (a) two-thirds of $\pi\pi$ decays are to $\pi^+\pi^-$ and one-third to
$\pi^0\pi^0$ and (b) there are equal numbers of decays to $K^+K^-$ and $K^0\bar K^0$. 
  Here the errors are the statistical and the systematic
errors added in
quadrature, and for the systematic error, the common parts related
to the
MDC tracking, kinematic fitting,
efficiency of
the photon ID, and the number of $\psi(2S)$ events cancel.

There is a strong $f_0(980)f_0(2200)$ with a signal significance of
$7.1\sigma$ in the $\chi_{c0}\to\pi^+\pi^-K^+K^-$ decay. The mass and
width of the $f_0(2200)$ are optimized as M = $2170 \pm
20^{+10}_{-15}$ MeV/c$^2$ and $\Gamma = 220 \pm 60^{+40}_{-45}$ MeV/c$^2$, and
$${\cal B}[\chi_{c0}\to f_0(980)f_0(2200)] {\cal
B}[f_0(980)\to\pi^+\pi^-] {\cal B}[f_0(2200)\to K^+K^-] =
(8.42\pm1.42^{+1.65}_{-2.29})\times 10^{-4}.$$  Changing
the spin-parity of the $f_0(2200)$ in the fit or adding an additional
resonance, shows that the spin-parity of the $f_0(2200)$ is well determined and no additional resonance is needed in the $f_0(2200)$
mass region. However, compared to the nearby states $f_0(2100)$ and
$f_2(2150)$, its properties are still less well known \cite{PDG}, and more
experimental data are needed.

Another significant decay mode to $f_0(1370)f_0(1710)$ is also found
with a significance of
$6.5\sigma$, where $f_0(1370)$ decays to $\pi^+\pi^-$ and $f_0(1710)$
decays to $K^+K^-$. The fitted mass and width of the
$f_0(1710)$ are M $=1760\pm15^{+15}_{-10}$ MeV/c$^2$ and $\Gamma = 125\pm
25^{+10}_{-15}$~MeV/c$^2$. The fitted mass is somewhat higher than the PDG value
\cite{PDG}. The spin 0 component can be separated from spin 2 clearly,
and replacing $f_0(1710)$ with either of the tensors $\rho(1450)$ or  
$\rho(1700)$ does not give a reasonable fit to the data.
A free fit to $f_0(1370)$ gives a fitted mass of $1265 \pm
30^{+20}_{-35}$ MeV/c$^2$ 
and
a width of $350 \pm 100^{+105}_{-60}$ MeV/c$^2$. The
corresponding branching fraction is $${\cal B}[\chi_{c0}\to
f_0(1370)f_0(1710)] {\cal B}[f_0(1370)\to\pi^+\pi^-] {\cal B}[f_0(1710)\to K^+K^-] =
(7.12\pm1.85^{+3.28}_{-1.68})\times 10^{-4}.$$

Besides the intermediate $(\pi^+\pi^-)(K^+K^-)$ decay modes listed in
Table I, we tried the following combinations in the fit: $f_0(1370)f_0(1370)$, $f_0(1370)f_0(1500)$, $f_0(1500)f_0(1370)$, $f_0(1500)f_0(1500)$, and $f_0(1500)f_0(1710)$. None of them improved the log
likelihood more than 5. So we didn't include these processes in the
final solution of our fit and set upper limits at the 90\% C.L.:
\begin {eqnarray*}{\cal B}[\chi_{c0}\to f_0(1370)f_0(1370)] {\cal B}
[f_0(1370)\to\pi^+\pi^-] {\cal B} [f_01370)\to K^+K^-] < 2.9\times 10^{-4},\\
{\cal B}[\chi_{c0}\to f_0(1370)f_0(1500)] {\cal B}
[f_0(1370)\to\pi^+\pi] {\cal B} [f_0(1500)\to K^+K^-] < 1.8\times 10^{-4},\\
{\cal B}[\chi_{c0}\to f_0(1500)f_0(1370)] {\cal B}
[f_0(1500)\to\pi^+\pi^-] {\cal B} [f_01370)\to K^+K^-] < 1.4\times 10^{-4},\\
{\cal B}[\chi_{c0}\to f_0(1500)f_0(1500)] {\cal B}
[f_0(1500)\to\pi^+\pi^-] {\cal B} [f_0(1500)\to K^+K^-] < 0.55\times 10^{-4},\\
{\cal B}[\chi_{c0}\to f_0(1500)f_0(1710)] {\cal B}
[f_0(1500)\to\pi^+\pi^-] {\cal B} [f_0(1710)\to K^+K^-] < 0.73\times 10^{-4}.
\end {eqnarray*}

From the results of the fit, the branching
fraction ${\cal B}[\chi_{c0}\to K^*(892)^0$
$\bar
K^*(892)^0\to\pi^+\pi^- K^+K^-] = (8.09\pm 1.69^{+2.29}_{-1.99})\times
10^{-4}$ is obtained. Using the branching fraction
of $K^*(892)^0$
to the charged $K\pi$ mode, which is taken as $
{2}\over{3}$, we get
\begin {eqnarray*}
{\cal B}[\psi(2S)\to\gamma\chi_{c0}\to\gamma K^*(892)^0\bar
K^*(892)^0] = (1.68\pm 0.35^{+0.47}_{-0.40})\times 10^{-4},\\{\cal B}[\chi_{c0}\to K^*(892)^0\bar
K^*(892)^0] = (1.82\pm 0.38^{+0.52}_{-0.45})\times 10^{-3}.\end{eqnarray*}
The values obtained here are consistent with those in
Ref. \cite{kk}.

Most of the peak around a $K\pi$ mass $\sim 1430$ MeV/c$^2$ is fitted with
$K^*_0(1430)\bar
K^*_0(1430)$ and $K^*_0(1430)\bar
K^*_2(1430)$
$+~c.c.$, with only a small contribution from
$K^*_2(1430)\bar
K^*_2(1430)$. The measured branching fractions are \begin {eqnarray*}{\cal
B}[\chi_{c0}\to K^*_0(1430)\bar
K^*_0(1430)^0\to\pi^+\pi^- K^+K^-] = (10.44\pm
2.37^{+3.05}_{-1.90})\times
10^{-4},\\{\cal
B}[\chi_{c0}\to K^*_0(1430)\bar
K^*_2(1430)^0 + c.c. \to\pi^+\pi^- K^+K^-] = (8.49\pm
1.66^{+1.32}_{-1.99})\times
10^{-4}.\end{eqnarray*} A free fit to
$K^*_0(1430)$ gives a fitted mass of $1455 \pm 20\pm15$ MeV/c$^2$ and
a width of $270 \pm 45^{+30}_{-35}$ MeV/c$^2$. The fitted number of
$K^*_2(1430)\bar
K^*_2(1430)$ events is about 11\% of that of $K^*_0(1430)\bar
K^*_0(1430)$, and the signal significance of $K^*_2(1430)\bar
K^*_2(1430)$ mode is about $3.3\sigma$. These measurements are important
for the study of $\chi_{c0}$ decays, as well as those of
$K^*_0(1430)$ and $K^*_2(1430)$.

Using the results shown in Table IV and the branching fractions of
$K_1(1270)\to K\rho(770)$ and $K_1(1400)\to K^*(892)\pi$ given by the PDG
\cite{PDG}, we determine
\begin {eqnarray*}{\cal B}[\chi_{c0}\to K_1(1270)^{+}K^{-} + c.c.] =
(6.66\pm1.31^{+1.60}_{-1.51})\times 10^{-3},\\{\cal B}[\chi_{c0}\to
K_1(1400)^{+}K^{-} + c.c.] <
2.85\times 10^{-3},\end {eqnarray*}
at the 90\% C.L..
To accommodate this, a mixing angle of $\theta >
57^{\circ}$ is required. A flavor-SU(3)-violating
$K_1(1270)-K_1(1400)$ asymmetry is observed. The asymmetries with
opposite character for
$\psi(2S)$ and $J/\psi$ decays were also observed in the BES\,I data
\cite{AV}, where for $\psi(2S)$ data, $\theta < 29^{\circ}$ and for
$J/\psi$ data $\theta > 48^{\circ}$.

\section{SUMMARY}

In summary, a partial wave analysis on
$\chi_{c0}\to\pi^+\pi^-K^+K^-$ in $\psi(2S)\to\gamma\chi_{c0}$ decay
is performed using a
 sample of
14 million $\psi(2S)$ events.
From the
fit we find significant contributions to the $\chi_{c0}$ decays from
the channels  $f_0(980)f_0(980)$,
$f_0(980)f_0(2200)$, $f_0(1370)f_0(1710)$,
$K^*(892)^0\bar K^*(892)^0$, $K^*_0(1430)\bar K^*_0(1430)$,
$K^*_0(1430)\bar K^*_2(1430) + c.c.$ and $K_1(1270)K$. The mass and width of the $f_0(1710)$ are determined to be $1760\pm15^{+15}_{-10}$ MeV$/c^2$ and $125\pm25^{+10}_{-15}$ MeV$/c^2$, and those of the $f_0(2200)$ are $2170\pm20^{+10}_{-15}$ MeV$/c^2$ and $220\pm60^{+40}_{-45}$ MeV$/c^2.$
Flavor-SU(3)-violating $K_1(1270)-K_1(1400)$ asymmetry is
observed, with the mixing angle $\theta > 57^{\circ}$. 

\section{Acknowledgment}
 We wish to thank Qiang Zhao for useful suggestions. The BES collaboration thanks the staff of BEPC for their hard
efforts. This work is supported in part by the National Natural
Science Foundation of China under contracts Nos. 10491300,
10225524, 10225525, 10425523, the Chinese Academy of Sciences under
contract No. KJ 95T-03, the 100 Talents Program of CAS under
Contract Nos. U-11, U-24, U-25, and the Knowledge Innovation
Project of CAS under Contract Nos. U-602, U-34 (IHEP), the
National Natural Science Foundation of China under Contract No.
10225522 (Tsinghua University), and the Department of Energy under
Contract No. DE-FG02-04ER41291 (University of Hawaii).

\end{document}